\documentclass[letterpaper]{article}
\usepackage{aaai23} 
\usepackage{times} 
\usepackage{helvet} 
\usepackage{courier} 
\usepackage[hyphens]{url} 
\usepackage{graphicx} 
\usepackage{graphicx} 
\usepackage{natbib} 
\usepackage{caption} 
\usepackage{mathtools}
\usepackage{amssymb}
\usepackage{nicefrac}
\usepackage{subcaption}
\usepackage{wrapfig}
\newcommand{\qizwo}{{\sc QI$^2$}}
\newcommand{\qir}{{\sc QI$^2$R}}
\newcommand{\mlqi}{{\sc MLQI$^2$}}
\newcommand{\blqi}{{\sc BLQI$^2$}}
\newcommand{\hlqi}{{\sc HLQI$^2$}}
\newcommand{\shlqi}{{\sc SHLQI$^2$}}

\title{\qizwo{} - an Interactive Tool for Data Quality Assurance}
\author{
    Simon Geerkens\textsuperscript{\rm 1},
    Christian Sieberichs\textsuperscript{\rm 1},
    Alexander Braun\textsuperscript{\rm{1}},
    Thomas Waschulzik\textsuperscript{\rm 2}}

\affiliations {
    \textsuperscript{\rm 1}University of Applied Sciences Düsseldorf, 40476 Düsseldorf, Germany,\\
    \textsuperscript{\rm 2}Siemens Mobility GmbH, 91058 Erlangen, Germany\\
    \{simon.geerkens, christian.sieberichs, alexander.braun\}@hs-duesseldorf.de, thomas.waschulzik@siemens.com
}
\begin{document}
\maketitle
\begin{abstract}
\begin{quote}
The importance of high data quality is increasing with the growing impact and distribution of ML systems and big data. Also the planned AI Act from the European commission defines challenging legal requirements for data quality especially for the market introduction of safety relevant ML systems. In this paper we introduce a novel approach that supports the data quality assurance process of multiple data quality aspects. This approach enables the verification of quantitative data quality requirements. The concept and benefits are introduced and explained on small example data sets. How the method is applied is demonstrated on the well known MNIST data set based an handwritten digits. 
\end{quote}
\end{abstract}

\section{Introduction}
Up to now ML systems are mostly used in areas, where wrong decisions only have minor consequences. The increasing performance of ML compared to classical systems especially for perceptions tasks in natural complex environments gives hope for e.g. driver-less vehicle operation in foreseeable future. The hopes of using ML also for perception tasks in areas of higher risk currently may not be fulfilled, due to the required trustworthiness of the systems including assured high quality of the used data sets may not be proven. To give a legal framework for the application of ML systems the European AI act \cite{european_comission_laying_2021} is currently under development. Simultaneously there are multiple projects from research and industry working on the topic of ML Systems in high risk areas like KI-Absicherung \cite{fingscheidt_safety_2022} \cite{habli_integrated_2021} and safe.trAIn \cite{siemens_safetrain_2022}. High risk ML systems have to fulfill the requirements according to \cite{european_comission_laying_2021} Chapter 2 "REQUIREMENTS FOR HIGH-RISK AI SYSTEMS" Article 10 "Data and data governance" Point 3: 
\begin{quote}
    Training, validation and testing data sets shall be relevant, representative, free of errors and complete.
\end{quote} 

In this paper we present a novel approach to visually represent and quantify data quality, and demonstrate this approach with several different application examples. The aim is to provide new tools to understand and manage the data quality. The approach is a part of the QUEEN-method (\textbf{Qu}alitätsgesicherte \textbf{e}ffiziente \textbf{E}ntwicklung vorwärtsgerichteter künstlicher \textbf{N}euronaler Netze, \textit{quality-assured efficient development of neural networks}) \cite{waschulzik_qualitatsgesicherte_1999} a high-level procedure for quality assured development of neural networks. QUEEN gives guidance for developing explainable artificial intelligence (XAI) based on the development of dedicated prepossessing. It is supported with quality indicators that are used as key performance indicators for the achieved complexity reduction for the given task. Quality assuring the processing of every single step before the data is used as input for a neural network supports the trustworthiness for the data bases and is combined with lightweight AI systems that can be interpreted in detail. QUEEN defines two data quality assurance methods called \qizwo{} (\textbf{i}ntegrated \textbf{q}uality \textbf{i}ndicator) \cite{geerkens_anwendung_2021} and ECS (\textbf{e}quivalent \textbf{c}lasses \textbf{s}ets) \cite{sieberichs_anwendung_2021}. In this paper we want to show the mathematical bases as well as the applications of \qizwo{} and its underlying methods in terms of data quality assurance. The method quantifies the complexity of input-output relationship defined with a data set. The term complexity herein is representative for a measurement of the amount of non-linearities in input-output relationships. It is based on the pairwise measurement of input and output distances what makes it attractive for a wide variety of applications due to simply adapting the distance metric. The ECS is not part of this paper but is covered in the submission \cite{sieberichs_ecs_2023}.

The \qizwo{} and its respective methods are a handy tool in terms of quality assurance due to a low dimensional representation of data. It quantifies neighborhood input-output relationship behaviors over a set of data points. High dimensional anomalous structures and relationships in data are represented by anomalous structure in a simple visualisation. It is possible to directly and accurately address several quality aspects - e.g. global linearity, outlier, simple and difficult sub-structures, clustering structure, inconsistent data, discontinuities - by simply interacting with its visual representation.

\section{Related Work}
Data quality in general is a widely researched topic due to the growing necessity and usage of data in our every day life. But there are still a lot of problems to be tackled. In \cite{wang_beyond_1996} Data quality is defined with respect to the intended use of the consumer. Although data should be intrinsically good they state that data quality needs to be a context dependant term in order to be appropriate for the desired task. The term of data quality is furthermore split into many dimensions like accuracy, consistency, completeness, safety, timeliness etc. in \cite{sidi_data_2012}. \cite{gualo_data_2021} considers the ISO/IEC 25012, an accepted industry standard for data quality and gives a model based approach. In addition to that \cite{pipino_data_2002} states a separation into subjective and objective quality assessment with three functional forms. The diversity of the definition of data quality has to be resolved in the future to give a guidance to implement the European AI Act. A consensus seems to exist, that data quality has intrinsic properties - accuracy, completeness, consistency, etc - as well as properties that are more dependent on use - timeliness, presentation quality etc.

As a first glance in data quality assurance descriptive statistics \cite{holcomb_fundamentals_2016} is used. Herein statistical methods for measuring central tendencies, dispersion and association are used for basic understanding of variables in data. The most common displaying method for those statistics are visualisations via scatter plots and histograms. Our proposed method extends descriptive statistical methods with a new visualization that gives deeper inside to the spatial relationships in the typically high dimensional data to support dedicated quality assurance and enables direct interactions between different visualisations of the data.

Dimensional reduction methods like PCA \cite{jolliffe_principal_1990}, tSNE \cite{maaten_visualizing_2008} or UMAP \cite{mcinnes_umap_2020} are a further visually representative quality assurance. These methods map the high dimensional data into low dimensions that may be easier analysed and understood by humans. Nevertheless they suffer from a high loss of information and might impose misleading interpretations. The herein proposed method of \qizwo{} is a human interpretable visual representation of the data according to neighborhood relationships. It therefore represents high dimensional information of the data that are necessary for efficient quality assurance. It does not contain the data itself in the visualisation but it is computed on the original values and the structural behavior.

Further metrics claiming approaches for general data quality assurance try to cover as many dimensions of data quality as possible by testing data against predefined rules and assumptions. In contrast to descriptive statistics those methods do not define data quality due to a statistical and distributional structure that is met by the overall data. They assess data in order to meet certain rules and assumptions. \cite{iannone_pointblank_2022} published the pointblank R package for agent based data quality assurance. This package contains quality assessments on a higher level basis with testing values in specific columns or rows against defined functions. One can test if the data is correct in terms of the values either being greater, lower, equal or between certain values or follow structural or informational rules defined by the user. 
Another general quality assurance method is DEEQU published in \cite{schelter_automating_2018} and \cite{schelter_deequ_2018}. This package allows the user to define specific assumption based unit tests for data at large scale. Similar to the previous one this package also adds row and column based checks to a pipeline for data quality assurance. In addition to unit tests on a single data set there is the possibility of anomaly detection over time. 
A further assumption based approach for data quality assurance is shown in \cite{heinrich_assessing_2018}. The probability-based method herein outputs a value representing the probability of a data set being free from internal errors with respect to given rules. In comparison to the previous rule based approaches this method does not apply rules row or column based. The rules defined for this assurance are more like relationships between individual columns of data sets representing correlations. 
Despite a more complex application of rule based checks all mentioned methods are heavily assumption and rule based and therefore require high knowledge of the data set and accurate assumptions. Assumptions on data although will never allow full coverage of quality dimensions because one can not know what one does not know. Due to a representative visualisation of input-output relationship behavior our proposed method does not need any assumptions for certain quality aspects. It can be used without any knowledge about the data and does not need thresholds or possible coincidences between data dimensions.

For deeper quality assessment more specific methods need to be taken into account. Herein methods developed for a specific purpose will be the key. In the field of outlier detection for example there exist a variety of different approaches all tackling the same problem. A first strategy is using density-based outlier detection algorithm \cite{breunig_lof_2000}. As a first glance this method outputs a value similar to a probability determining if a data point being an outlier. This value is computed based on how locally isolated the data point is. For this the local density of each point is compared to the local density of their $k$-nearest neighbors using the local reachability density (LRD). Another clustering based approach uses DBSCAN \cite{ester_density-based_1996} as clustering algorithm and repeats a specific clustering and cluster merging algorithm to a state in which every cluster has its own $\epsilon$ value regarding DBSCAN. After that $minPts$ value of DBSCAN will be computed for each cluster with respect to the smallest $minPts$ of that cluster. Regarding $minPts$ clusters will be classified as anomalous \cite{tran_manh_thang_anomaly_2011}.

In further development density based clustering was used as a preprocessing in outlier detection combined with inter cluster distance based classification of clusters as anomalous. Methodologically similar approaches differing in clustering, distance computation and anomaly detection are \cite{fawzy_outliers_2013} \cite{samara_enhanced_2022} \cite{samara_complete_2023}. The first one is using the fixed width clustering algorithm for finding clusters in a data set followed by computing inter-cluster distances. Clusters are classified as anomalous by looking at the average inter-cluster distance and its deviation from the mean inter-cluster distance. The second one above is using the previously mentioned DBSCAN clustering algorithm in combination with a anomalous classification based on inverse distance weighting (IDW). The third one is using OPTICS \cite{ankerst_optics_1999} algorithm for clustering and an inter cluster distance based anomalous classification with kringing methods. The methods of \qizwo{} could as well be seen as a clustering-based analysis method but differs from the mentioned methods due to the fact that outlier detection is not based on thresholds and clustering as preprocessing directly. Clustering as preprocessing can lead to loss in information regarding individual data points. Furthermore the proposed method gives a wider variety in quality assurance than only tackling one specific problem.

\section{\qizwo{} - The Quality Indicator}
Since the algorithm presented visually represents the relationships between input and output changes within a data set $P$, there must first be a division of the data set into input space and output space. Therefore every data point can be described by 
\begin{equation}
    \vec{p^T} \coloneqq \begin{pmatrix}
    vi_{1} \hdots vi_{I} vo_{1} \hdots vo_{O} 
\end{pmatrix}
\end{equation}
with $I$ values of the input-space $vi_{i} \epsilon \mathbb{R}$ and $O$ values of the output-space $vo_{o} \epsilon \mathbb{R}$. 

With this separation the method of \qizwo{} can be applied. First of all it is necessary to calculate the pairwise distances regarding input-space and output-space throughout every possible pair $P^2 \coloneqq \{x \coloneqq (\vec{p_{1}}, \vec{p_{2}}) \mid \vec{p_{1}}, \vec{p_{2}} \epsilon \mathbb{R}^{(I+O)}\}$ of data points. For this every possible metric $M$ can be applied. For example euclidean or cosine distance for scalar values and SSIM \cite{wang_image_2004} or mutual information \cite{kanade_image_2004} for images. Despite using a certain metric for the input-space the output-space can have a different metric applied. Take as an example a quality assurance of an image classification data set. In the input-space (images) the SSIM can be applied to compare the structural information as distance metric but in the output-space (classes as scalar values) the euclidean distance can be applied.
\begin{equation}
    d_{RI}(x) \coloneqq M(x)
\end{equation}
\begin{equation}
    d_{RO}(x) \coloneqq M(x)
\end{equation}

After that the distances will be normalized by mean for comparability. In the following this will be shown exemplary for the input-space.
\begin{equation} \label{norm_dist}
    d_{NRI}(x) \coloneqq \nicefrac{d_{RI}(x)}{\frac{\sum_{y \epsilon P^2} d_{RI}(y)}{|P^2|}}
\end{equation}

Having the pairwise normalized distances in input- and output-space it is possible to compute a first representative value for a data set described by the data points taken into account. This value describes the complexity and quality of this data set by comparing the relationship between output changes following on certain input changes.
\begin{equation} \label{QI2}
    QI^2R(P) \coloneqq \frac{1}{|P^2|} \sum_{x \epsilon P^2} (d_{NRI} - d_{NRO})^2
\end{equation}
This representative value gives a first shot in quality assurance of a data set by describing the non-linearity. The higher the value, the more non-linear parts are in the data set. A second handy property of this value is that it describes a two dimensional data set with random normal distribution with a value of one. If the \qir{} of a data set is higher than one, the input-output behaviore of the data set is more complex than a random normal distribution. This occurs e.g. if parts of the data set are more likely to have more linear relationships between input and output but are separated by a sudden change in direction of the gradient.

The next step in quality assurance of a data set is a local complexity representation. Therefore the \qir{} can be repeatedly computed over subsets of the original data set. For this the data set needs to be sub sampled by e.g. building subsets based on increasing k-nearest neighborhoods around every data point. This means for every data point in the data set there will be a number of subsets depending on how much neighbors should be taken into account. Another possibility is to divide the data set into subsets with respect to a high dimensional sphere around each data point with increasing diameter. With those subsets it is now possible to compute a matrix of local \qizwo{} (\mlqi{}).
\begin{equation}
    mlqi^2_{i,k}(P) \coloneqq QI^2R(KNN(P, p_{i}, k))
\end{equation}
$KNN(P, p_{i}, k)$ represents the buildup of subsets based on $k$-nearest neighbors around the respective data point $p_{i}$ in the data set $P$. The computation of \mlqi{} considers every pairwise distance between possible pairs of points in the subset.

Visualizing a matrix human interpretable is hardly possible in this case. Thus a visualization via a histogram of local \qizwo{} (\hlqi{}) was chosen. 
\begin{equation}
    hlqi^2_{v,k}(P) \coloneqq \sum_{i=1}^{|P|}{I^3({mlqi^2}_{i,k}(P), v)\cdot{blqi^2}_{i,k}(P)}
\end{equation}

The term $blqi^2_{i,k}$ (boolean matrix of local \qizwo{}, \blqi{}) is responsible for preventing boundary effects by comparing neighborhoods. It checks whether the \qir{} for the current neighborhood has already been computed, since this neighbourhood already exists on the basis of another data point. If it has already been computed this calculation will be dropped in the computation of $hlqi^2_{v,k}$.
\begin{equation}
    blqi^2_{i,k}(P) \coloneqq \begin{cases}
        1, & \text{for }\forall{j<i} \\
        & | KNN(P,p_{j},k) \neq KNN(P,p_{i},k)\\
        0, & \text{else}
        \end{cases}
\end{equation}

$I^3(h,v)$ is responsible for sorting the \mlqi{} into the desired histogram by checking if the receiving value $h$ lies in the bin $[v|v+1)$.
\begin{equation}
    I^3(h,v) \coloneqq \begin{cases}
        1, & \text{for } v \le \frac{h-min_{hi}}{binsize_{hi}} < v+1 \\
        0, & \text{else}
        \end{cases}
\end{equation}

As a last step the histogram is scaled to its maximum value in every $k$ so that it contains relative values for the bins and it is gamma calibrated due to visual reasons for lower values (scaled histogram of local \qizwo{}, \shlqi{}). 
\begin{equation}
    shlqi^2_{v,k}(P) \coloneqq \left(\frac{hlqi^2_{i,k}(P)}{\sum_{s=1}^{|P|}{hlqi^2_{s,k}(P)}}\right)^{gamma_{hi}}
\end{equation}

\subsection{Benefits of \qizwo{}}
\begin{figure}
    \centering
    \includegraphics[width=\linewidth]{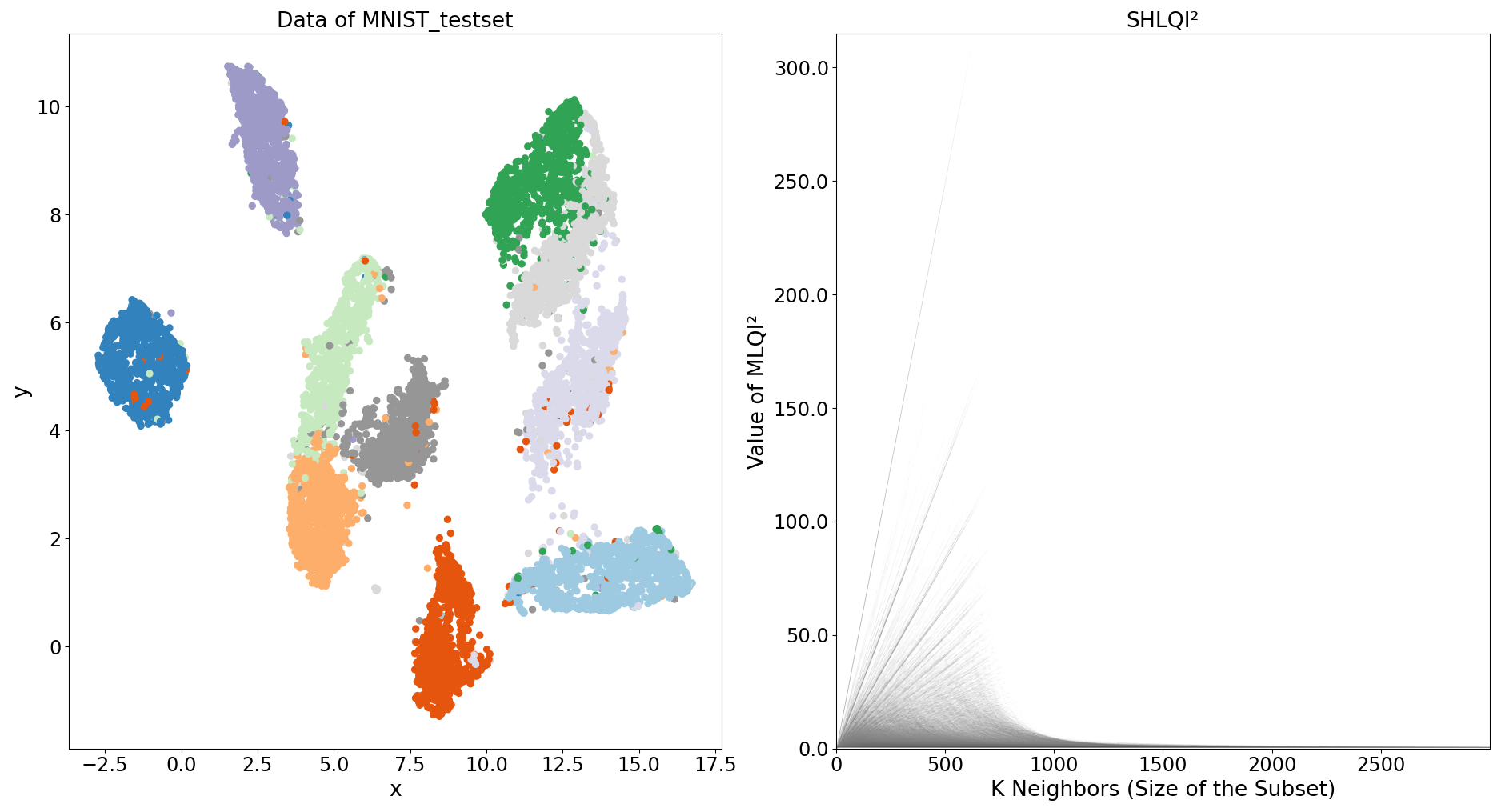}
    \caption{UMAP and \shlqi{} representation of the MNIST handwritten digits test data set}
    \label{fig:MNIST_test}
\end{figure}

The most important advantage of this method is a compressed and representative visualization of local input-output relationship behavior through various different subsets based on every data point. By this the visualization and knowledge of the computational structure can be used to exactly identify certain interesting data points and quality aspects simply by interaction between the histogram and the data set. Those aspects are e.g. outliers, linear subtasks, discontinuities and many more. As stated above in comparison to quality assurance with a dimension reduction algorithm this method gives a more accurate representation due to no possible distortion in data relationships. This is simply because of directly considering the original values for every data point as it is represented in the data set in the \shlqi{} plot. Nevertheless a dimension reduction algorithm can be used to visualize the data set human interpretable. In combination with the interaction between the \shlqi{}and the data visualisation it is possible to get a deep dive into the structure and input-output relation of the data set. Furthermore the \shlqi{}can be used for quantitative quality assurances due to analysis based on quantitative requirements. An Example of a visualization of \shlqi{} and a dimensional reduced two dimensional visual representation of MNIST \cite{lecun-mnisthandwrittendigit-2010} test set is given in figure \ref{fig:MNIST_test}. 

\subsection{Using \qizwo{} to understand Data Quality} 
The methods of \qizwo{} can be used to determine certain quality aspects in data sets as stated in the introduction. In order to assure quality two separate analyses are possible
\begin{itemize}
    \item global analysis
    \item local analysis
\end{itemize}
In a first glance of a global analysis the value of \qir{} as well as its density function and the cumulative distribution can be used to determine a global complexity. Those methods are calculated based on either the whole data set or based on the sum of increasing clusters around each data point. For further local data quality assurance the \shlqi{} can be used. The most important aspect in quality assurance with \shlqi{} is that the visualisation can be used as a human interpretable representation of a high dimensional data set and therefore has to be treated like that. It is necessary to watch out for anomalous and differing areas in the \shlqi{} from the 'general' structure of the histogram. Since it analyses the input-output behavior of a data set by representative visualization it detects anomalous or interesting structures with anomalous visualisation. Interesting or anomalous structures in data can be both positive and negative in terms of quality and need a separate treatment afterwards. Nevertheless there are certain characteristics due to the computation of the \mlqi{} that are not straight forward and therefore differ in the visualization of \shlqi. For this differentiation data sets need to be classified regarding their task, approximation or classification due to huge differences in structure and analysing methods between these two tasks. \\

\begin{figure*}
    \centering
    \begin{subfigure}{0.49\textwidth}
        \includegraphics[width=\textwidth]{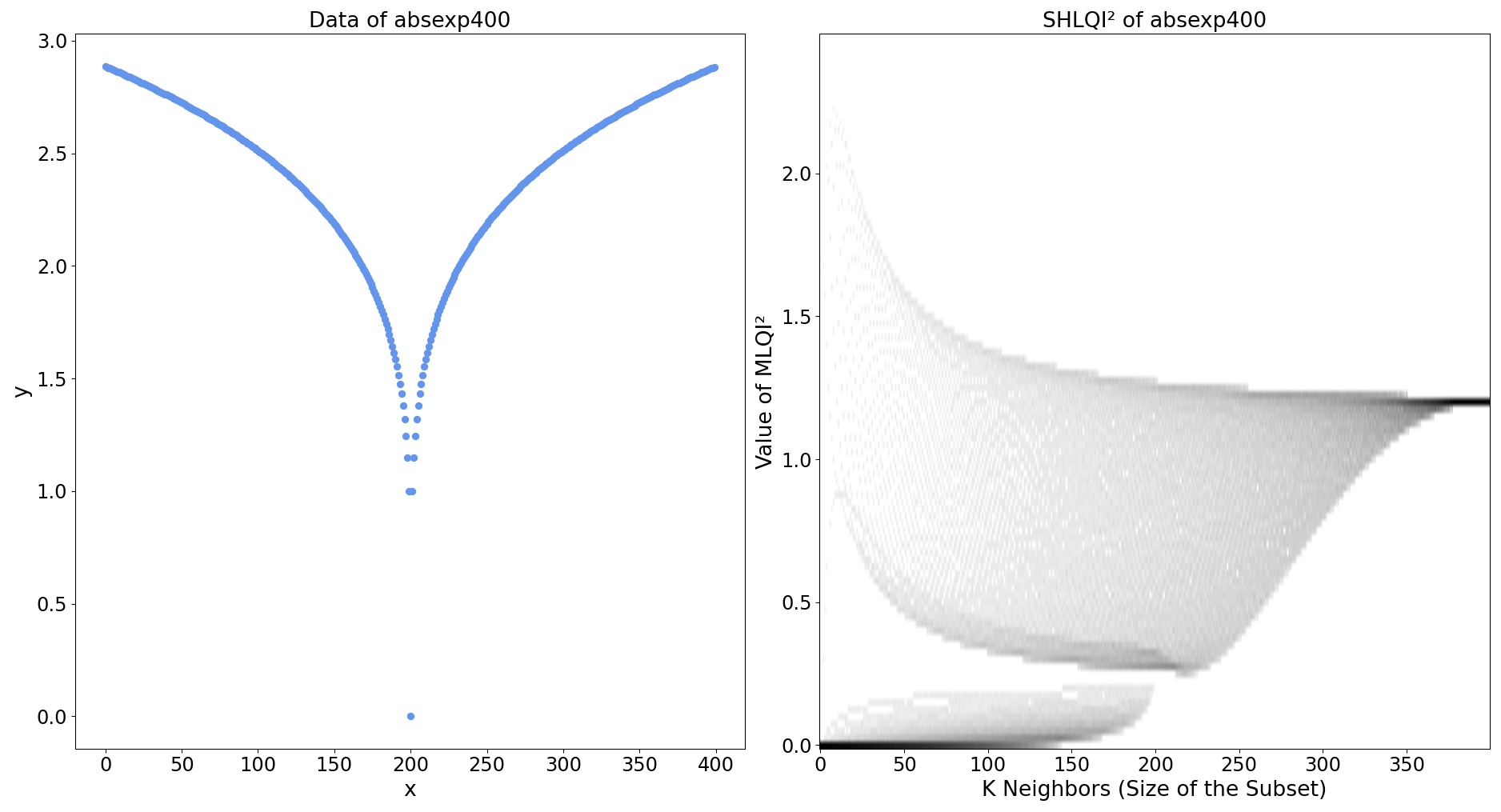}
        \caption{\shlqi{} for function $f(x) \coloneqq |x-200|^{0.2}$}
        \label{fig:absexp_example}
    \end{subfigure}
    \hfill
    \begin{subfigure}{0.49\textwidth}
        \includegraphics[width=\textwidth]{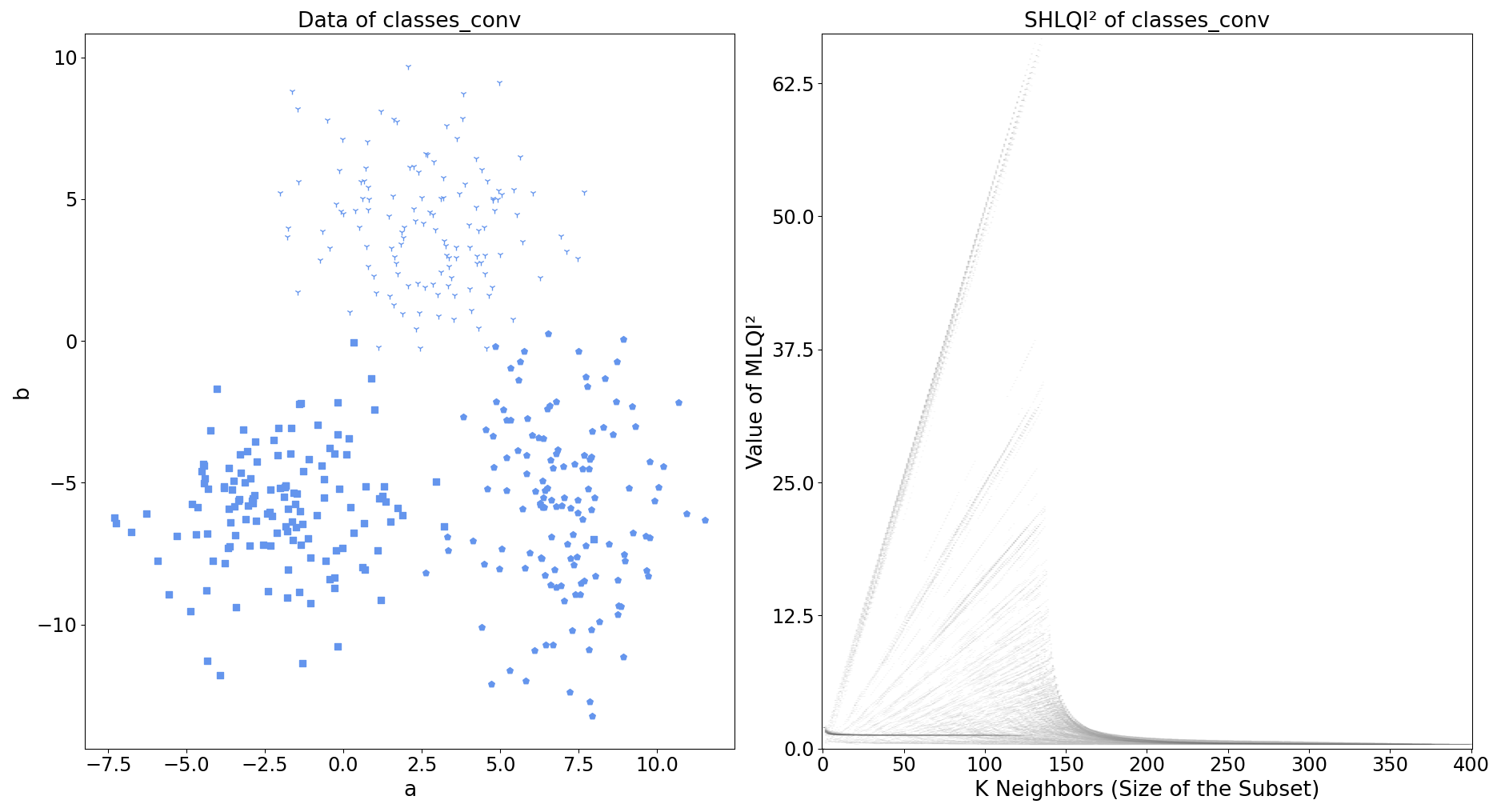}
        \caption{\shlqi{} for a classification}
        \label{fig:cluster}
    \end{subfigure}
    \caption{Comparison of \shlqi{} of approximation and classification task}
\end{figure*}

\textbf{Approximation tasks}. For this kind of task the histogram will be build up between bins $mlqi^2_{i,k} \epsilon [0|2.5]$ in most cases. It has a wider coverage of different bins in the local area and will decrease in coverage towards one single value at maximum k. There are several different characteristics of this histogram that can be interpreted directly. As mentioned above the final value of the \shlqi{} at maximum $k$ describes the overall complexity of the whole data set. This is a good starting point in terms of quality assessment due to the fact that for every dimensionality of the input data space there is a specific value for a randomly normal distributed data set. As mentioned above for one dimension in input space it is the value one. For increasing dimension the value decreases due to the curse of dimensionality. If the final value of the \shlqi{} at maximum $k$ is higher than this, the overall data set has a more complex input-output relationship than a randomly normal distribution. In the case of figure \ref{fig:absexp_example} this is the case due to the sudden change in direction at $x=200$. 
A second handy feature of the \shlqi{} are the higher and lower values in local areas. If there are many representative chunks in higher bins in local areas the data set has locally complex areas that need to be treated externally. In contrast to that many chunks in lower bins means many local areas that have a more or less linear relationship between input and output. Holes in the histogram represent highly complex areas in the data set like sparsely covered input space, sudden changes in data, outlier and so on.\\

\textbf{Classification tasks}. For classification tasks the structure of the \shlqi{} is completely different from the one for approximation tasks. First of all its general structure is similar throughout different data sets. By this one can directly differentiate between a classification and approximation task just by looking on the histogram. The structure of the \shlqi{} for classification tasks follows strict rules due to the computation based on characteristics of those tasks. The most specific characteristic are steep rises of complexity towards values way higher than $2.5$ in combination with sudden drops of those rises at a data set specific location. Those steep rises and sudden drops are caused by the normalized distance computation (equation \ref{norm_dist}) in the output space. In terms of a classification the distances between same classes is zero due to the definition of distance metrics. If the computation gets to a point, where a data point with a different class is the next neighbor within the subset there is a sudden change in output distances. According to the computational structure considering every pairwise distance there are now a few distances $d_{RI}\neq 0$. Lets take a representative example. Let the subset $P$ around $p_i$ be a set of 100 data points. within this set there are 99 points with the same class as $p_i$ and one point with a different class. Therefore this data point with a different class has a pairwise distance to every other point differing from zeroe.g. $d_{RO}(x)=1$. Now $|P^2|=100 \cdot 100=10.000$ possible data point pairs. Every pairwise distance $d_{RO}$ will be divided by $\frac{\sum_{y\epsilon |p^2|}{d_{RO}(y)}}{|P^2|} = \frac{100}{10.000} = 0.01$ according to equation \ref{norm_dist}. In comparison to the values of $d_{NRI}$ the result of equation \ref{QI2} will be a high value due to $d_{NRO}(y) >> d_{NRI}(y)$ for all pairs $y$ for which the two data points have two different classes. The sudden drop in values of SHLQI² is caused by a further data point with a different class from the major class interferring with the subset by incrementing the neighborhood. By this the denominator of equation \ref{norm_dist} is now nearly doubled due to double the amount of pairwise distances $d_{RO} \neq 0$ compared to a slight rise in the amount of possible data point pairs. \\
Another characteristic of the \shlqi{} for classification tasks is a specific area for local k within the bins $mlqi^2_{i,k} \epsilon (1|2)$. Herein most complexities are located in a $e^{-x}$-like function representing homogeneous groups of classes. The smaller the value of this noticeable are, the denser the clusters. By adding a weak numerical stabiliser in the denominator in equation \ref{norm_dist} in order to make it numerical stable in terms of comparing identical values the input-output relationship of homogeneous cluster behave somewhat similar to a random distribution. The behavior can be described as randomly distributed inputs referring to one identical output. The darker this area, the more subsets are local homogenous clusters representing a single class.

\section{Exemplary Data Quality Assurance}
As stated before the \shlqi{} can be used to determine certain quality aspects in data sets with respect to a local analysis. Certain aspects need special treatments regarding the identification via interaction with \shlqi{}. We want to present some exemplary identifications with examplary and state-of-the-art data sets. \\

\begin{figure*}
    \centering
    \begin{subfigure}{0.49\textwidth}
        \includegraphics[width=\textwidth]{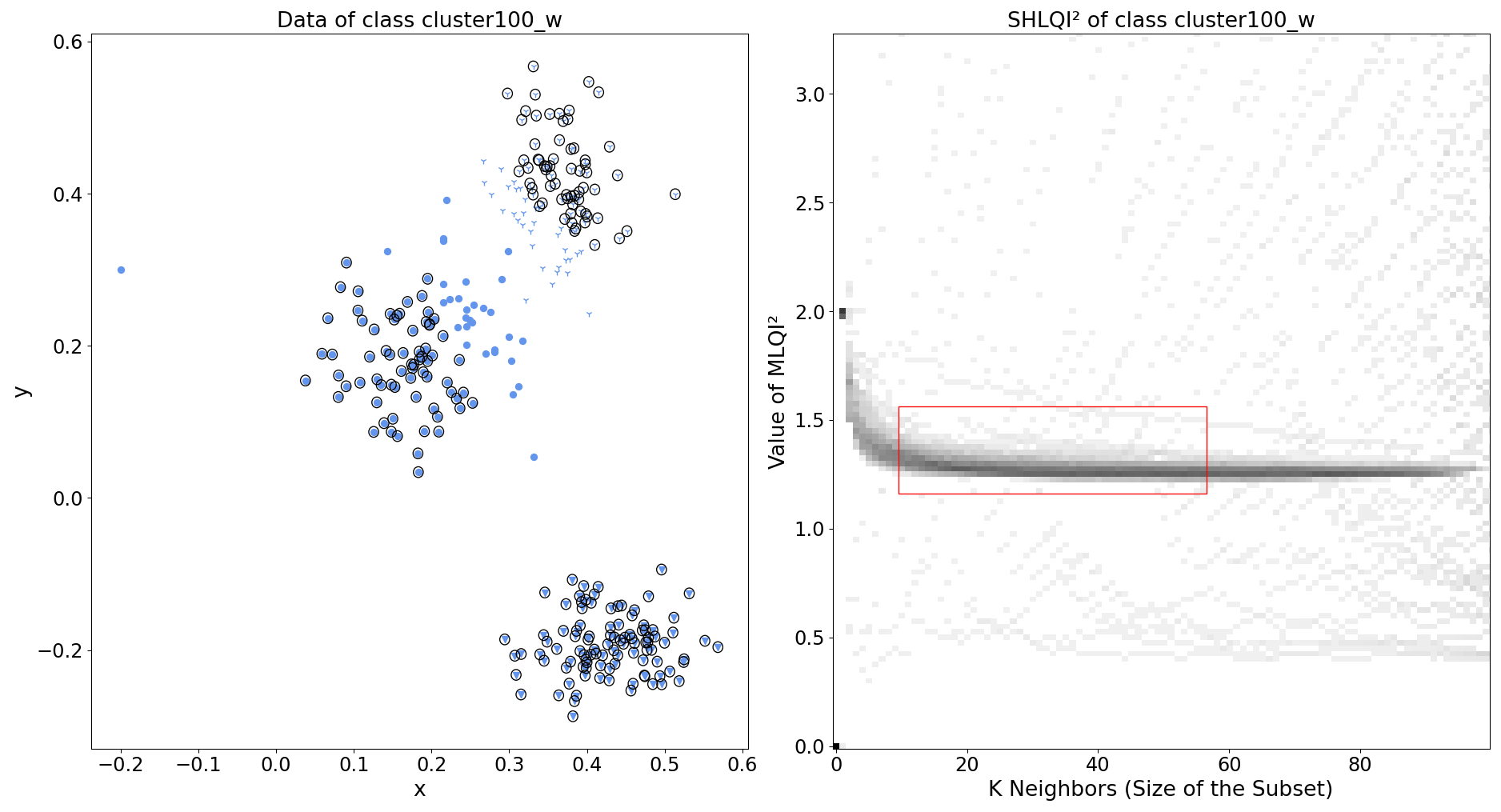}
        \caption{Identification of data points in homogeneous groups}
        \label{fig:homogeneous}
    \end{subfigure}
    \hfill
    \begin{subfigure}{0.49\textwidth}
        \includegraphics[width=\textwidth]{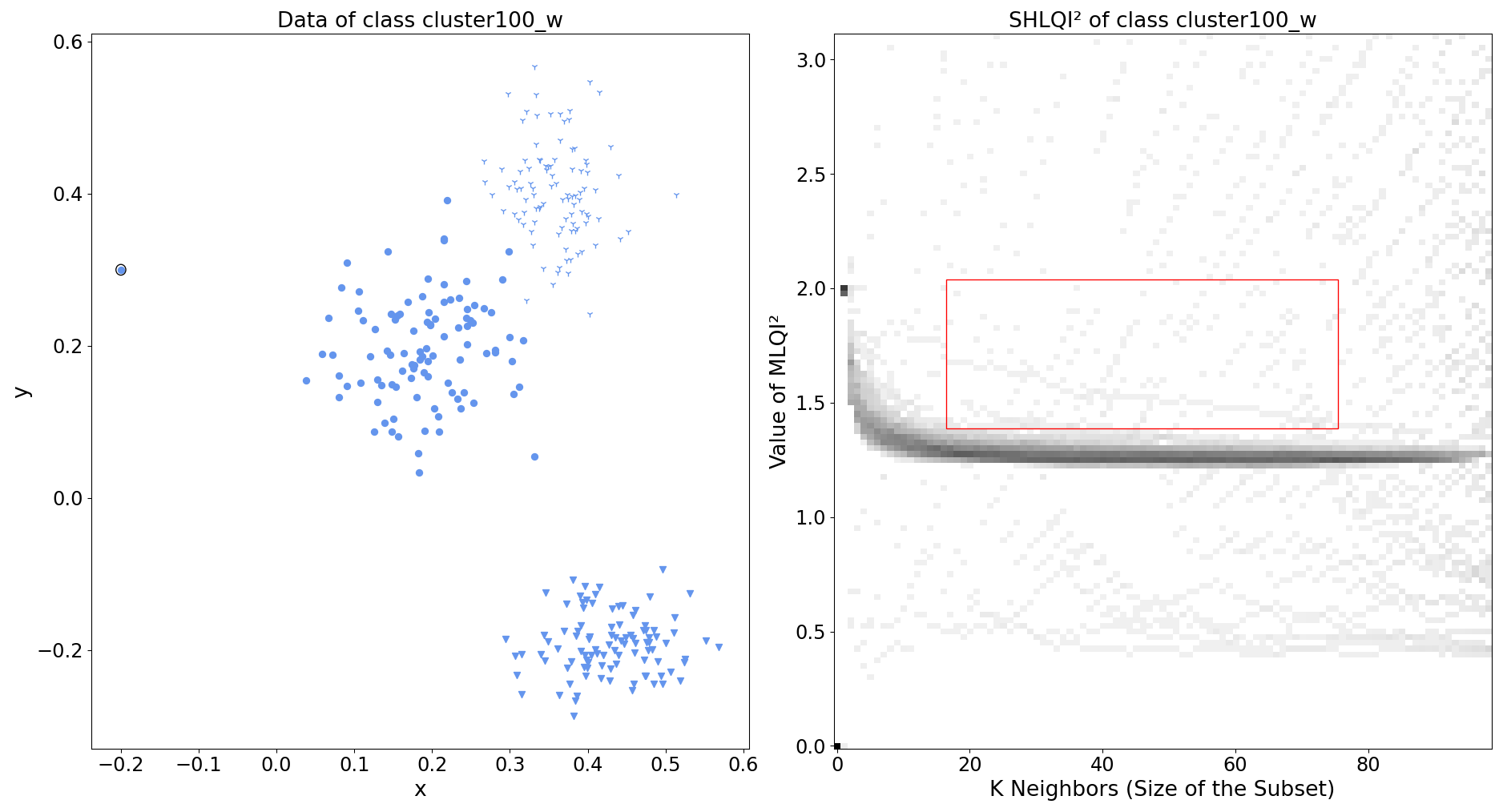}
        \caption{Identification of out of distribution data points}
        \label{fig:OoD}
    \end{subfigure}
    \caption{Identification of different structural and individual distribution aspect}
\end{figure*}
\begin{figure}
    \centering
    \includegraphics[width=\linewidth]{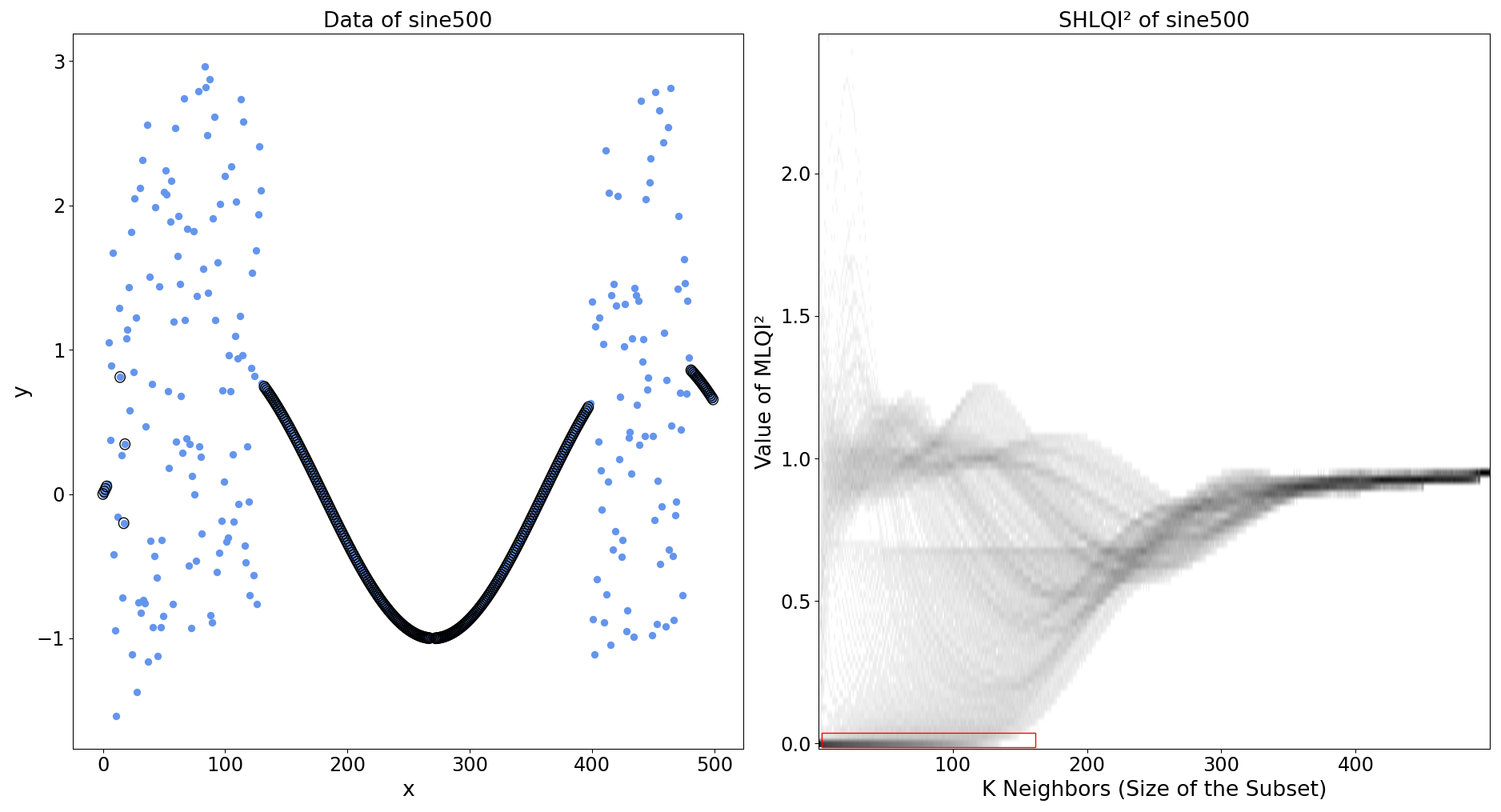}
    \caption{Identification of simple subset}
    \label{fig:simple_subset}
\end{figure}

\textbf{Structural and individual distribution}. The overall distribution has a huge impact in data quality. The \shlqi{} can be used in order to determine the overall structure due to homogeneity of clusters as mentioned above. As an example figure \ref{fig:homogeneous} visualizes the identification of homogeneous clusters. Herein a classification data set was analysed due to its data points being in a homogeneous cluster in which every data point has 60 direct nearest neighbors with the same class. The identification therefore needs to be a check if for a certain range of $k$ the values of \shlqi{} for a point is not leaving the characteristic area of homogeneous clusters.

In contrast to analysing a data set in terms of clustering it is possible to identify out of distributional data points individually. Out of distribution therefore needs to be defined as a data point being outside of its respective classes cluster but not interfering with another class. While interfering with another class this data point would be a misclassificational outlier. The identification is quite similar to the one of homogeneous clusters. The only difference is, that due to average normalization of the input distances the value of \qir{} will be higher within a defined boundary of {$mlqi^2_{i,k} \epsilon [1|2]$. This is caused by considering significant higher normalized distances between the out of distribution data point and every other point inside the respective cluster in equation \ref{QI2}. Therfore this identification is a check if for a certain local range of $k$ the values of \shlqi{} for a point is within the range slighlty above the homogeneous characteristic and below two as presented in figure \ref{fig:OoD}.\\

\textbf{Locally simple input-output relationship}. Data sets with multiple thousands of data points can be very complex when it comes to interpretation of the whole data set. Its input-output relationship behavior can have quite non-trivial changes or interruptions due to nonlinearities in the real world. The data points can and will have heavily complex relationships especially with increasing input dimensions. Therefore it is nice if a data set has a subset that can be specially treated due to a rather incomplex input-output relationship behavior. For those points one can apply the divide-and-conquer principle. The simpler subset can be used to get further interpretability and stability into the system the data is applied on. For example an ML system can generate high robustness against inputs near or inside this sub structure of the data. As an exemplary visualisation of identification of simple subset, an artificially created data set representing a simple sine wave with heavy noise due to undefined reasons was created. The noisy parts follow no specific rules in terms of input-output correlation. In contrast to that, the sine is following a more strict relationship between input and output dimensions. The more linear this relationship is, the lower is the local complexity for the data points in those parts. A simple subset therefore will be visible due to persistent and falling low complexities in local areas. The longer the complexities are in a low area the more data points form a simple subset. Visually this case is shown in figure \ref{fig:simple_subset}.

It can be seen clearly that the marked lower complexities in the \shlqi{} directly refer to the sine based parts of the data set. Those data points are following defined simpler structural rules and can be treated like a simple subtask.\\

\section{MNIST training data quality assurance}
To explain the use of the new method the popular MNIST handwritten digits training data set has been chosen. The data set consists of 60.000 different 28x28 gray scale images. We applied the computation of \shlqi{} onto the whole data set with a maximum neighborhood of 3000 neighbors. At this point the expectation was that nearly the whole histogram will be covered by the previous mentioned steep rises with extremely high value due to an average amount of 6000 data points per class. According to this amount of points per class we expected the set to have a pretty decent clustering and therefore homogeneous groups with couple hundreds or thousands of neighbors from the same class. The next data point with another class would then cause an explosion of the value of $mlqi^2_{i,k}$. As shown in figure \ref{fig:MNISTfull} this assumption is totally met. The complexities of MNIST reached pretty high values with a peak at $mlqi^2_{i,k}\approx 1100$ as well as high coverage of the histogram with classification characteristic steep rises in complexities.\\

\begin{figure}
    \centering
    \includegraphics[width=\linewidth]{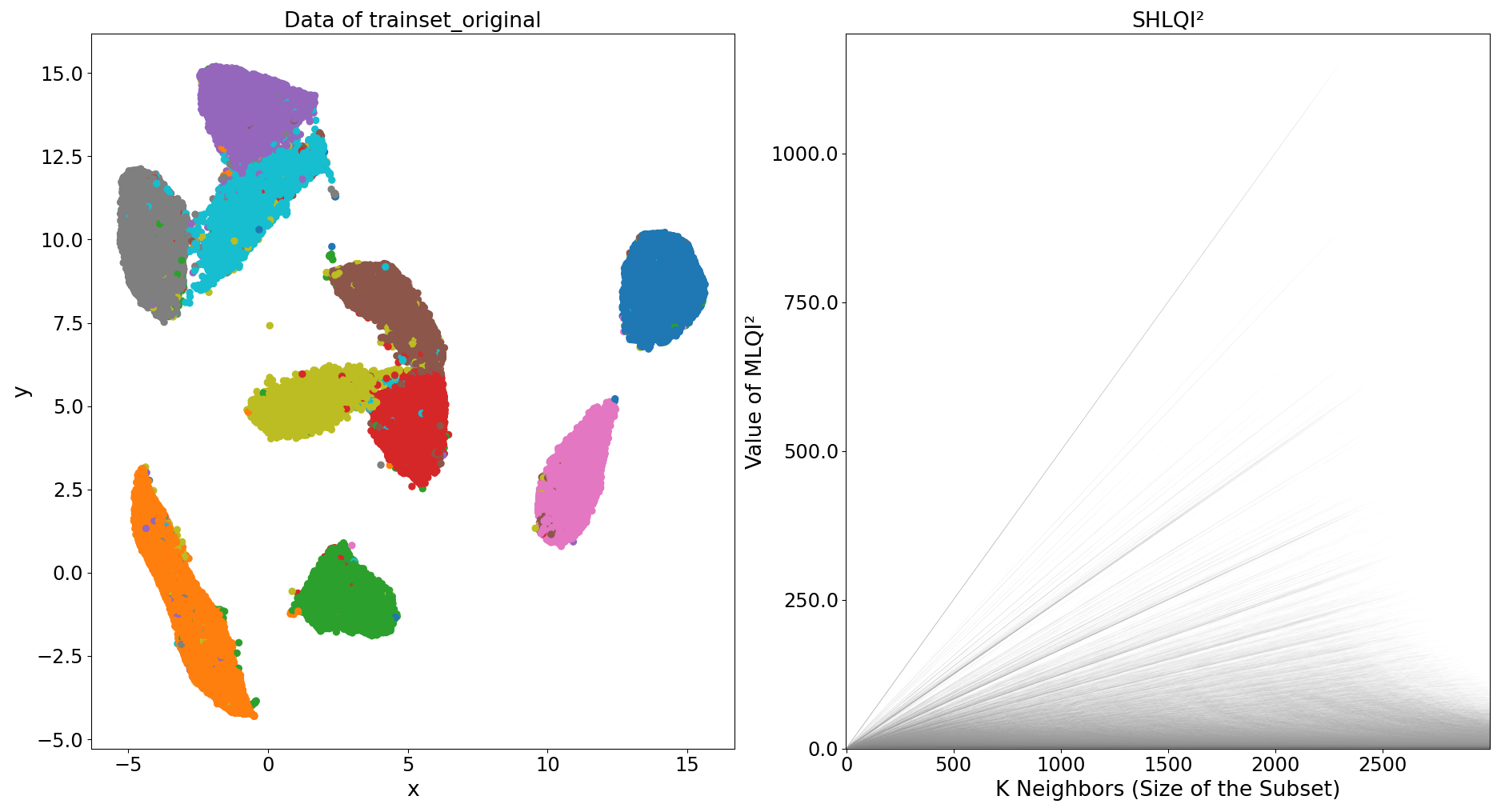}
    \caption{\shlqi{} for MNIST training data with euclidean distance}
    \label{fig:MNISTfull}
\end{figure}

\textbf{Structural distribution}. After first overall quality assurance a deeper analysis in order to determine the structural distribution can be done. In order to get a good knowledge of how well the data is homogeneously clustered the proposed identification of structural distribution was applied. Figure \ref{fig:homogeneousMNIST} shows the correlation between the proposed identification and its respective black marked data points. 
\begin{wraptable}{r}{.3\linewidth}
    \centering
    \begin{tabular}{c|c}
         $k$ & $p_{cluster}$ \\[0.5ex] 
         \hline \\
         100 & 25405 \\
         200 & 16769 \\
         300 & 12743 \\
         500 & 8230 \\
         1000 & 2919
    \end{tabular}
    \caption{Data points in homogeneous clusters due to neighborhood size with identical class}
    \label{tab:cluster}
\end{wraptable}
Interesting to see is that for the classes "one" (bottom left cluster), "zero" (rightmost cluster) and "two" (left cluster below "zero") almost every data point is part of a homogeneous cluster with respect to 300 nearest neighbors. In contrast to that the cluster "nine" has only two representative data points. This already shows the difference in quality assurance between \shlqi{} and dimension reduction algorithms. While UMAP represents a qualitative homogeneous cluster for class "nine", the \shlqi{} shows that it is build up by quantitative rather small homogeneous clusters within local areas. The UMAP visualisation is misleading in terms of overall structural distribution. Furthermore table \ref{tab:cluster} shows the amount of data points building homogeneous cluster $p_{cluster}$ with at least $k$ nearest neighbors having an identical class.

\begin{figure}
    \centering
    \includegraphics[width=\linewidth]{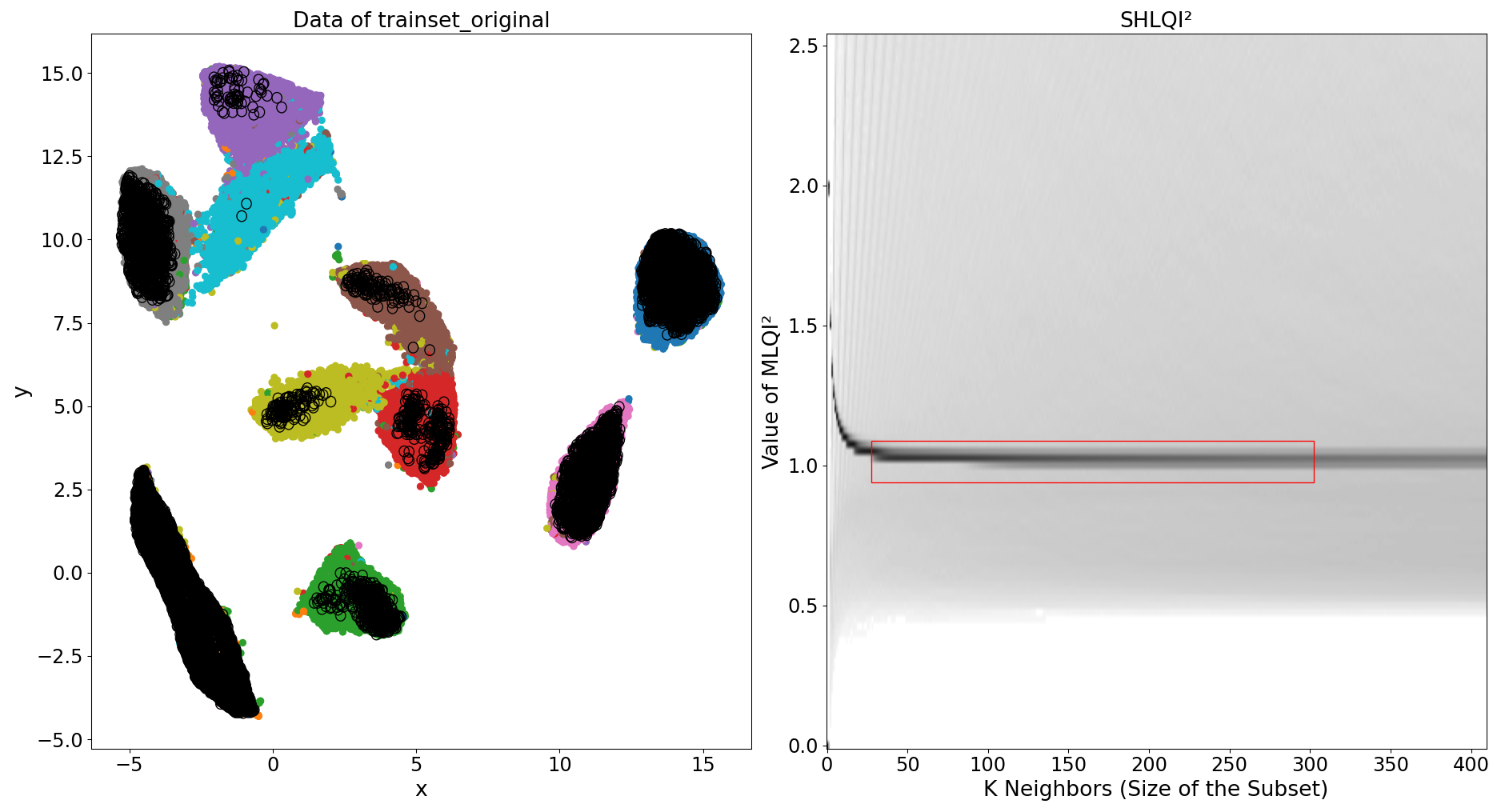}
    \caption{Homogeneous clustering in MNIST training data with a neighborhood size of 300 nearest neighbors with identical class}
    \label{fig:homogeneousMNIST}
\end{figure}

\textbf{Outlier detection}. An outlier in classification tasks can be detected due to low \mlqi{} at $k=1$ and a steep rise in complexity for local areas $k\epsilon [5|25]$. A low complexity at $k=1$ represents data points having one direct neighbor with another class. This might as well be the case in terms of classification boundaries but in combination with steep complexity rises in local areas these classification boundaries will be filtered out. The filtering happens because of an evenly distributed amount of data points with another and the same class in local areas for boundaries. This would result in lower complexity values. As stated in above classification tasks have steep rises in complexity due to one or a few examples being in an homogeneous cluster which is the case for outliers. Examples for outliers, their class according to the data and their detection in MNIST training data regarding euclidean distance are shown in figure \ref{fig:MNIST_outliers}.\\
\begin{figure}
    \centering
    \begin{subfigure}{0.5\linewidth}
        \includegraphics[width=\textwidth, height=\textwidth]{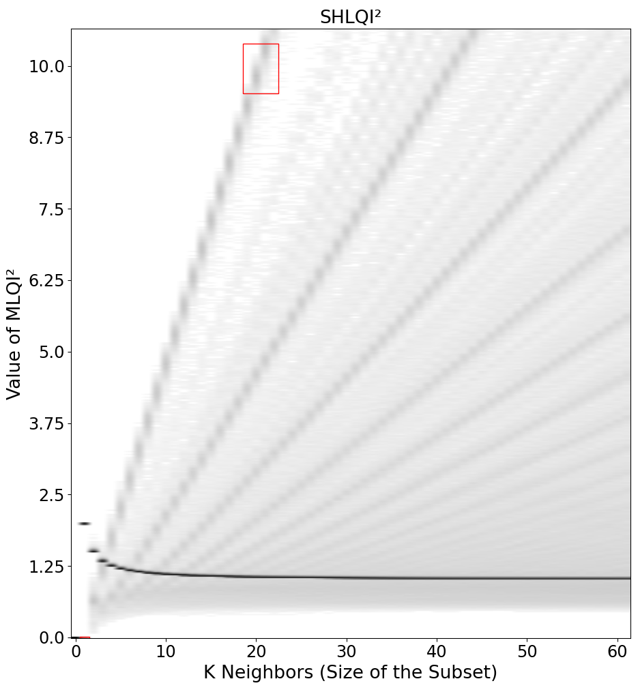}
        \caption{Identification of outliers}
        \label{fig:outlier_markings}
    \end{subfigure}
    \hfill
    \begin{subfigure}{0.47\linewidth}
        \raisebox{.13\textwidth}{
            \includegraphics[width=1\textwidth, height=0.7\textwidth]{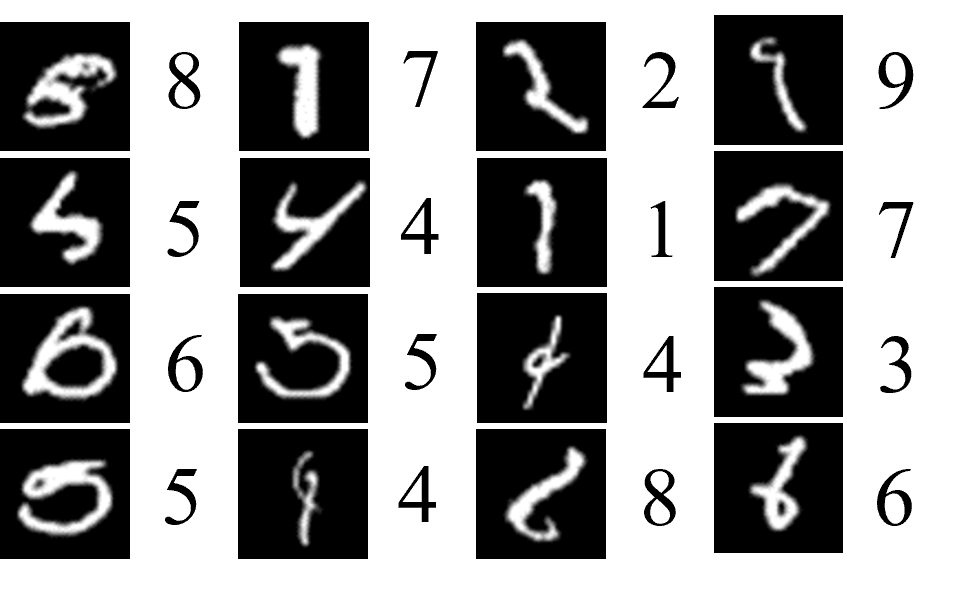}
        }
        \caption{Examples of detected outliers}
        \label{fig:outliers}
    \end{subfigure}
    \caption{Outlier detection in MNIST data set}
    \label{fig:MNIST_outliers}
\end{figure}

\section{Conclusion}
In this paper we presented a novel quality assurance method based on visually representing the neighborhood relationship. We gave concrete examples for assurance of defined quality aspects and tested them exemplary with the MNIST training data. The \qizwo{} can be used to get deeper structural and individual insights regarding qualitative and quantitative aspects of the data. The new tool-set based on \qizwo{} enables quantitative requirements for the quality assurance process. As an example a requirement for the quality assurance of the data set may be formulated, that for all potential outliers identified with \shlqi{} in neighbourhoods of 10 to 50 examples the output of every example has to be validated. With this measure the probability of wrong class labels will be reduced significantly with relative low costs for quality assurance activities. The range for which neighbourhoods the validation is required may be e.g. defined according to the required safety integrity level of the application.
Especially in combination with a representative visualisation of the data this tool has many handy features for an overall efficient interactive quality assurance process.

\bibliography{Geerkens}

\begin{thebibliography}{30}
\providecommand{\natexlab}[1]{#1}

\bibitem[{Ankerst et~al.(1999)Ankerst, Breunig, Kriegel, and
  Sander}]{ankerst_optics_1999}
Ankerst, M.; Breunig, M.~M.; Kriegel, H.-P.; and Sander, J. 1999.
\newblock {OPTICS}: {Ordering} {Points} to {Identify} the {Clustering}
  {Structure}.
\newblock In \emph{Proceedings of the 1999 {ACM} {SIGMOD} {International}
  {Conference} on {Management} of {Data}}, {SIGMOD} '99, 49--60. New York, NY,
  USA: Association for Computing Machinery.
\newblock ISBN 1-58113-084-8.
\newblock Event-place: Philadelphia, Pennsylvania, USA.

\bibitem[{Breunig et~al.(2000)Breunig, Kriegel, Ng, and
  Sander}]{breunig_lof_2000}
Breunig, M.~M.; Kriegel, H.-P.; Ng, R.~T.; and Sander, J. 2000.
\newblock {LOF}: {Identifying} {Density}-{Based} {Local} {Outliers}.
\newblock In \emph{Proceedings of the 2000 {ACM} {SIGMOD} {International}
  {Conference} on {Management} of {Data}}, {SIGMOD} '00, 93--104. New York, NY,
  USA: Association for Computing Machinery.
\newblock ISBN 1-58113-217-4.
\newblock Event-place: Dallas, Texas, USA.

\bibitem[{Burton et~al.(2022)Burton, Hellert, Hüger, Mock, and
  Rohatschek}]{fingscheidt_safety_2022}
Burton, S.; Hellert, C.; Hüger, F.; Mock, M.; and Rohatschek, A. 2022.
\newblock Safety {Assurance} of {Machine} {Learning} for {Perception}
  {Functions}.
\newblock In Fingscheidt, T.; Gottschalk, H.; and Houben, S., eds., \emph{Deep
  {Neural} {Networks} and {Data} for {Automated} {Driving}}, 335--358. Cham:
  Springer International Publishing.
\newblock ISBN 978-3-031-01232-7 978-3-031-01233-4.

\bibitem[{Deng(2012)}]{lecun-mnisthandwrittendigit-2010}
Deng, L. 2012.
\newblock The MNIST Database of Handwritten Digit Images for Machine Learning
  Research [Best of the Web].
\newblock \emph{IEEE Signal Processing Magazine}, 29(6): 141--142.

\bibitem[{Ester et~al.(1996)Ester, Kriegel, Sander, and
  Xu}]{ester_density-based_1996}
Ester, M.; Kriegel, H.-P.; Sander, J.; and Xu, X. 1996.
\newblock A {Density}-{Based} {Algorithm} for {Discovering} {Clusters} in
  {Large} {Spatial} {Databases} with {Noise}.
\newblock In \emph{Proceedings of the {Second} {International} {Conference} on
  {Knowledge} {Discovery} and {Data} {Mining}}, {KDD}'96, 226--231. AAAI Press.
\newblock Event-place: Portland, Oregon.

\bibitem[{{European Comission}(2021)}]{european_comission_laying_2021}
{European Comission}. 2021.
\newblock {LAYING} {DOWN} {HARMONISED} {RULES} {ON} {ARTIFICIAL} {INTELLIGENCE}
  ({ARTIFICIAL} {INTELLIGENCE} {ACT}) {AND} {AMENDING} {CERTAIN} {UNION}
  {LEGISLATIVE} {ACTS}.

\bibitem[{Fawzy, Mokhtar, and Hegazy(2013)}]{fawzy_outliers_2013}
Fawzy, A.; Mokhtar, H. M.~O.; and Hegazy, O. 2013.
\newblock Outliers detection and classification in wireless sensor networks.
\newblock \emph{Egyptian Informatics Journal}, 14(2): 157--164.

\bibitem[{Geerkens(2021)}]{geerkens_anwendung_2021}
Geerkens, S. 2021.
\newblock Anwendung und {Validierung} des {SHLQI}² auf realen {Beispielmengen}
  und neuronale {Netzwerke}.

\bibitem[{Gualo et~al.(2021)Gualo, Rodriguez, Verdugo, Caballero, and
  Piattini}]{gualo_data_2021}
Gualo, F.; Rodriguez, M.; Verdugo, J.; Caballero, I.; and Piattini, M. 2021.
\newblock Data quality certification using {ISO}/{IEC} 25012: {Industrial}
  experiences.
\newblock \emph{Journal of Systems and Software}, 176: 110938.

\bibitem[{Heinrich et~al.(2018)Heinrich, Klier, Schiller, and
  Wagner}]{heinrich_assessing_2018}
Heinrich, B.; Klier, M.; Schiller, A.; and Wagner, G. 2018.
\newblock Assessing data quality – {A} probability-based metric for semantic
  consistency.
\newblock \emph{Decision Support Systems}, 110: 95--106.

\bibitem[{Holcomb(2016)}]{holcomb_fundamentals_2016}
Holcomb, Z. 2016.
\newblock \emph{Fundamentals of {Descriptive} {Statistics}}.
\newblock Routledge, 0 edition.
\newblock ISBN 978-1-351-97033-4.

\bibitem[{Iannone and Vargas(2022)}]{iannone_pointblank_2022}
Iannone, R.; and Vargas, M. 2022.
\newblock \emph{pointblank: {Data} {Validation} and {Organization} of
  {Metadata} for {Local} and {Remote} {Tables}}.
\newblock Https://rich-iannone.github.io/pointblank/,
  https://github.com/rich-iannone/pointblank.

\bibitem[{Jolliffe(1990)}]{jolliffe_principal_1990}
Jolliffe, I.~T. 1990.
\newblock {PRINCIPAL} {COMPONENT} {ANALYSIS}: {A} {BEGINNER}'{S} {GUIDE} - {I}.
  {Introduction} and application.
\newblock \emph{Weather}, 45(10): 375--382.

\bibitem[{Maaten and Hinton(2008)}]{maaten_visualizing_2008}
Maaten, L. v.~d.; and Hinton, G.~E. 2008.
\newblock Visualizing {Data} using t-{SNE}.
\newblock \emph{Journal of Machine Learning Research}, 9: 2579--2605.

\bibitem[{McInnes, Healy, and Melville(2020)}]{mcinnes_umap_2020}
McInnes, L.; Healy, J.; and Melville, J. 2020.
\newblock {UMAP}: {Uniform} {Manifold} {Approximation} and {Projection} for
  {Dimension} {Reduction}.
\newblock ArXiv:1802.03426 [cs, stat].

\bibitem[{Mock et~al.(2021)Mock, Scholz, Blank, Hüger, Rohatschek, Schwarz,
  and Stauner}]{habli_integrated_2021}
Mock, M.; Scholz, S.; Blank, F.; Hüger, F.; Rohatschek, A.; Schwarz, L.; and
  Stauner, T. 2021.
\newblock An {Integrated} {Approach} to a {Safety} {Argumentation} for
  {AI}-{Based} {Perception} {Functions} in {Automated} {Driving}.
\newblock In Habli, I.; Sujan, M.; Gerasimou, S.; Schoitsch, E.; and Bitsch,
  F., eds., \emph{Computer {Safety}, {Reliability}, and {Security}. {SAFECOMP}
  2021 {Workshops}}, volume 12853, 265--271. Cham: Springer International
  Publishing.
\newblock ISBN 978-3-030-83905-5 978-3-030-83906-2.
\newblock Series Title: Lecture Notes in Computer Science.

\bibitem[{Pipino, Lee, and Wang(2002)}]{pipino_data_2002}
Pipino, L.~L.; Lee, Y.~W.; and Wang, R.~Y. 2002.
\newblock Data {Quality} {Assessment}.
\newblock \emph{Commun. ACM}, 45(4): 211--218.
\newblock Place: New York, NY, USA Publisher: Association for Computing
  Machinery.

\bibitem[{Russakoff et~al.(2004)Russakoff, Tomasi, Rohlfing, and
  Maurer}]{kanade_image_2004}
Russakoff, D.~B.; Tomasi, C.; Rohlfing, T.; and Maurer, C.~R. 2004.
\newblock Image {Similarity} {Using} {Mutual} {Information} of {Regions}.
\newblock In Kanade, T.; Kittler, J.; Kleinberg, J.~M.; Mattern, F.; Mitchell,
  J.~C.; Nierstrasz, O.; Pandu~Rangan, C.; Steffen, B.; Sudan, M.; Terzopoulos,
  D.; Tygar, D.; Vardi, M.~Y.; Weikum, G.; Pajdla, T.; and Matas, J., eds.,
  \emph{Computer {Vision} - {ECCV} 2004}, volume 3023, 596--607. Berlin,
  Heidelberg: Springer Berlin Heidelberg.
\newblock ISBN 978-3-540-21982-8 978-3-540-24672-5.
\newblock Series Title: Lecture Notes in Computer Science.

\bibitem[{Samara et~al.(2022)Samara, Bennis, Abouaissa, and
  Lorenz}]{samara_enhanced_2022}
Samara, M.~A.; Bennis, I.; Abouaissa, A.; and Lorenz, P. 2022.
\newblock Enhanced efficient outlier detection and classification approach for
  {WSNs}.
\newblock \emph{Simulation Modelling Practice and Theory}, 120: 102618.

\bibitem[{Samara et~al.(2023)Samara, Bennis, Abouaissa, and
  Lorenz}]{samara_complete_2023}
Samara, M.~A.; Bennis, I.; Abouaissa, A.; and Lorenz, P. 2023.
\newblock Complete outlier detection and classification framework for {WSNs}
  based on {OPTICS}.
\newblock \emph{Journal of Network and Computer Applications}, 211: 103563.

\bibitem[{Schelter et~al.(2018{\natexlab{a}})Schelter, Lange, Schmidt, Celikel,
  Biessmann, and Grafberger}]{schelter_automating_2018}
Schelter, S.; Lange, D.; Schmidt, P.; Celikel, M.; Biessmann, F.; and
  Grafberger, A. 2018{\natexlab{a}}.
\newblock Automating large-scale data quality verification.
\newblock \emph{Proceedings of the VLDB Endowment}, 11(12): 1781--1794.

\bibitem[{Schelter et~al.(2018{\natexlab{b}})Schelter, Schmidt, Rukat,
  Kiessling, Taptunov, Biessmann, and Lange}]{schelter_deequ_2018}
Schelter, S.; Schmidt, P.; Rukat, T.; Kiessling, M.; Taptunov, A.; Biessmann,
  F.; and Lange, D. 2018{\natexlab{b}}.
\newblock {DEEQU} - {Data} quality validation for machine learning pipelines.
\newblock In \emph{{NeurIPS} 2018}.

\bibitem[{Sidi et~al.(2012)Sidi, Shariat~Panahy, Affendey, Jabar, Ibrahim, and
  Mustapha}]{sidi_data_2012}
Sidi, F.; Shariat~Panahy, P.~H.; Affendey, L.~S.; Jabar, M.~A.; Ibrahim, H.;
  and Mustapha, A. 2012.
\newblock Data quality: {A} survey of data quality dimensions.
\newblock In \emph{2012 {International} {Conference} on {Information}
  {Retrieval} \& {Knowledge} {Management}}, 300--304. Kuala Lumpur: IEEE.
\newblock ISBN 978-1-4673-1091-8 978-1-4673-1090-1.

\bibitem[{Sieberichs(2021)}]{sieberichs_anwendung_2021}
Sieberichs, C. 2021.
\newblock Anwendung und {Validierung} des {ECS} auf reale {Beispielmengen} und
  neuronale {Netzwerke}.

\bibitem[{Sieberichs et~al.(2023)Sieberichs, Geerkens, Braun, and
  Waschulzik}]{sieberichs_ecs_2023}
Sieberichs, C.; Geerkens, S.; Braun, A.; and Waschulzik, T. 2023.
\newblock {ECS} - an {Interactive} {Tool} for {Data} {Quality} {Assurance}.

\bibitem[{Siemens(2022)}]{siemens_safetrain_2022}
Siemens. 2022.
\newblock {safe.trAIn}.
\newblock \url{https://safetrain-project.de}.
\newblock Accessed: 2023-01-15.

\bibitem[{{Tran Manh Thang} and {Juntae
  Kim}(2011)}]{tran_manh_thang_anomaly_2011}
{Tran Manh Thang}; and {Juntae Kim}. 2011.
\newblock The {Anomaly} {Detection} by {Using} {DBSCAN} {Clustering} with
  {Multiple} {Parameters}.
\newblock In \emph{2011 {International} {Conference} on {Information} {Science}
  and {Applications}}, 1--5. Jeju Island: IEEE.
\newblock ISBN 978-1-4244-9222-0.

\bibitem[{Wang and Strong(1996)}]{wang_beyond_1996}
Wang, R.~Y.; and Strong, D.~M. 1996.
\newblock Beyond {Accuracy}: {What} {Data} {Quality} {Means} to {Data}
  {Consumers}.
\newblock \emph{Journal of Management Information Systems}, 12(4): 5--33.

\bibitem[{Wang et~al.(2004)Wang, Bovik, Sheikh, and
  Simoncelli}]{wang_image_2004}
Wang, Z.; Bovik, A.; Sheikh, H.; and Simoncelli, E. 2004.
\newblock Image {Quality} {Assessment}: {From} {Error} {Visibility} to
  {Structural} {Similarity}.
\newblock \emph{IEEE Transactions on Image Processing}, 13(4): 600--612.

\bibitem[{Waschulzik(1999)}]{waschulzik_qualitatsgesicherte_1999}
Waschulzik, T. 1999.
\newblock \emph{Qualitätsgesicherte effiziente {Entwicklung}
  vorwärtsgerichteter künstlicher {Neuronaler} {Netze} mit überwachtem
  {Lernen} ({QUEEN})}.
\newblock Ph.D. thesis, Technische Universität München, München.

\end{thebibliography}
\end{document}